\documentclass[12pt, conference, onecolumn]{IEEEtran}

\usepackage{graphicx}
\usepackage{tablefootnote}
\usepackage{environ}
\usepackage{bbm}
\usepackage{amsmath}
\usepackage{amsthm}
\usepackage{relsize}
\usepackage{csquotes}
\usepackage{algorithm, algorithmic}
\usepackage{mathtools}
\usepackage{nicefrac}
\usepackage{array}
\usepackage{cite}
\usepackage{geometry}
\allowdisplaybreaks

\usepackage{soul}
\setstcolor{red}

\newtheorem{lemma}{Lemma}

\newtheorem{theorem}{Theorem}

\DeclareMathOperator*{\argmin}{arg\,min}

\newtheorem{problem}{Problem}
\newcommand{\R}{\mathbbm{R}}

\newcommand{\av}{{\bf a}}
\newcommand{\bv}{{\bf b}}

\newcommand{\xv}{{\bf x}}

\newcommand{\zv}{{\bf z}}

\newcommand{\Ic}{{\cal I}}
\newcommand{\Jc}{{\cal J}}

\newcommand{\Nc}{{\cal N}}

\newcommand{\Wc}{{\cal W}}

\newcommand{\lambdav}{\hbox{\boldmath$\lambda$}}

\newcommand{\Thetav}{\hbox{\boldmath$\Theta$}}

\newcommand{\piv}{\hbox{\boldmath$\pi$}}
\newcommand{\rhov}{\hbox{\boldmath$\rho$}}

\newcommand{\<}{\left\langle}
\renewcommand{\>}{\right\rangle}

\newcommand{\ie}[0]{\textit{i.e.}, }

\newcommand{\eg}[0]{\textit{e.g.}, }

\newcommand{\vs}[0]{\textit{vs.}}

\NewEnviron{seq*}{
	\begin{equation*}
		\BODY
	\end{equation*}
	\normalsize	
}

\NewEnviron{sal}{
	\begin{align}
		\BODY
	\end{align}
	\normalsize	
}

\NewEnviron{seq}{
	\begin{equation}
		\BODY
	\end{equation}
	\normalsize	
}

\NewEnviron{myequation}{
	\begin{equation}
		\BODY
	\end{equation}
	\normalsize
}

\NewEnviron{myequation*}{
	\begin{equation*}
		\BODY
	\end{equation*}
	\normalsize	
}

\makeatletter
\def\endthebibliography{%
  \def\@noitemerr{\@latex@warning{Empty `thebibliography' environment}}%
  \endlist
}
\makeatother

\begin{document}
\title{Blind Optimal User Association  in \\ Small-Cell Networks 
}

\author{
\IEEEauthorblockN{Livia Elena Chatzieleftheriou}
\IEEEauthorblockA{Athens University of Economics \\ and Business, Greece\\
liviachatzi@aueb.gr}\\   
\IEEEauthorblockN{Georgios Paschos}
\IEEEauthorblockA{Amazon, Luxembourg\\
paschosg@amazon.com}\\
[0.9cm]  
\and
\IEEEauthorblockN{Apostolos Destounis}
\IEEEauthorblockA{Huawei Technologies, France\\
Apostolos.Destounis@huawei.com}\\\\
\IEEEauthorblockN{Iordanis Koutsopoulos}
\IEEEauthorblockA{Athens University of Economics \\ and Business, Greece\\
jordan@aub.gr}
}


\maketitle
\begingroup\renewcommand\thefootnote{}
\footnotetext{
To appear in IEEE International Conference on Computer Communications - INFOCOM,  10-13 May 2021, Virtual Conference. 
}
\endgroup

\begin{abstract}
We learn optimal user association policies for traffic from different locations to Access Points(APs), in the presence of unknown dynamic  traffic demand. 
We aim at minimizing a broad family of \(\alpha\)-fair cost functions that express various objectives in load assignment in the wireless downlink, such as total load or total delay minimization.
Finding an optimal user association policy in dynamic environments is challenging because traffic demand fluctuations over time are non-stationary and difficult to characterize statistically, which obstructs the computation of cost-efficient associations. 
Assuming arbitrary traffic patterns over time, we formulate the problem of online learning of optimal user association policies using the Online Convex Optimization (OCO) framework. 
We introduce a periodic benchmark for OCO problems that generalizes state-of-the-art benchmarks. 
We exploit inherent properties of the online user association problem and propose PerOnE, a simple online learning scheme that dynamically adapts the association policy to arbitrary traffic demand variations. We compare  PerOnE against our periodic benchmark and  prove that it enjoys the no-regret property, with additional sublinear dependence of the network size. To the best of our knowledge, this is the first work that introduces a periodic benchmark for OCO problems and a no-regret algorithm for the online user association problem.
Our theoretical findings are validated through results on a real-trace dataset. 
\end{abstract}

\vspace{-1mm}
\section{Introduction}\label{sec:intro}

Communication networks in the Beyond 5G (B5G)/6G era are envisioned to support ultra-low latency and bandwidth-damanding services, like those enabled by Internet of Things (IoT) or autonomous vehicles. 
Two key technological enablers of  such services in future communication networks are novel network architectures and the embedded use of Artificial Intelligence (AI) \cite{Strinati_2019_6G_Next_Frontier}. 
The new architectures will generalize the Coordinated MultiPoint transmission (CoMP), where APs cooperate to jointly serve requests within their coverage area, and each user's traffic may be served by more than one AP. 
The pervasive introduction of AI at the network edge, including distributed algorithms for proactive learning and prediction of unknown dynamic processes in the system, will enable the self-optimization of network resource allocation. 

In the envisioned ultra-dense wireless networks, devices will be in range of multiple Access Points (APs). These enhanced association possibilities will bring more degrees of freedom, and additional possibilities for optimization. 
The numerous devices and association alternatives call for a fast and agile user-to-AP association scheme. This is of vital importance for the upcoming bandwidth-demanding services, especially for the downlink, that supports the majority of traffic.  
Moreover, traffic demand at different locations heavily fluctuates during the day. This could happen, for example, 
due to sudden changes in the existing sources, or due to new unpredictable sources of traffic.  
Thus traffic is generally non-stationary during the day, which complicates its accurate statistical characterization and precludes the use of approaches that operate under stationary regimes, 
such as  Lyapunov optimization.

In this work we perform blind user associations on-the-fly, without any assumption or information about the actual traffic demand. We use the Online Convex Optimization (OCO) framework to produce updated solutions incrementally, by readjusting existing ones as new samples are observed.
We consider arbitrarily time-varying traffic demand for different locations and allocate it to APs, which we model as queues that capture their own load. 
These queues form an association policy whose cost belongs to a broad family of $\alpha$-fair functions of the load at APs, including as special cases several objectives, such as delay or load minimization. We aim at producing association policies that minimize regret, \ie the deviation of the cost of our online association, from that of the optimal offline association that knows in hindsight the traffic variations. 
We introduce OPS, a novel periodic benchmark that generalizes state-of-the-art,  and then propose PerOnE: an online algorithm that quickly adapts to unpredictable traffic variations, and that learns scalable and asymptotically optimal   user association policies for downlink traffic routing to different locations.

\subsection{Contributions}
The contributions of our work to the literature are as follows:
\begin{itemize}
    \item{ 
        We provide a model and an OCO formulation for the problem of Online Learning (OL) of how to dynamically associate traffic of geographic locations (and therefore users) to APs. 
        Our objective cost function models various targets for communication networks, such as AP load or delay  minimization. 
    }
    \item{
        We introduce Optimal Periodic Static (OPS), a novel peri- odic benchmark for OL problems that generalizes state-of-the-art. 
    In cases of traffic   periodicity, a benchmark where the association is the same during the day is not suitable. 
    OPS is appropriate to compare against, because the optimal policy will most likely be periodic as well.   
    }
    \item{We identify and exploit inherent properties of the online problem and design PerOnE, an efficient online algorithm that produces cost-effective association policies under arbitrary changes in traffic demand, with lack of information about the actual generated traffic and its statistical properties. PerOnE stems from Online Mirror Descend.     
    } 
    \item{We prove PerOnE's asymptotical optimality, as it achieves regret sublinear to the time-horizon 
    against OPS, that knows traffic variations in hindsight. Further,  PerOnE's regret also scales sublinearly with the network size, which renders it a valid association scheme for the upcoming large wireless networks. 
    } 
    \item Our evaluation  with publicly available traffic traces confirms the derived analytical results, showing that our algorithm achieves zero regret asymptotically. 
    In fact, its performance appears to be near-optimal with respect to a dynamic algorithm that chooses the optimum user association in each time slot. 
\end{itemize}

In section \ref{sec:related} we present the state-of-the-art. 
In section \ref{sec:model} we describe the model and the static user association problem.  
In section \ref{sec:regret} we introduce and analyse OPS. 
In section \ref{sec:algo} we perform a transformation of the 
static formulation concluding to an OCO formulation. We then design PerOnE, proving that its regret against OPS is sublinear both to the time horizon and to the problem dimension. 
Finally, in section \ref{sec:simulations} we evaluate our scheme on a real traffic dataset.

\section{Related Work}
\label{sec:related}

\textbf{User association (UA).} 
A widely adopted optimization framework is Network Utility Maximization (NUM) \cite{Kelly_1998_JoOR_networks_fairness_stability}, which is exemplified further for AP association.  
It considers a broad family of convex utility functions of the APs' load, capturing a variety of objectives, such as load balancing.  
The following works also consider convex cost functions.
In \cite{deVeciana_association} an iterative, distributed and deterministic UA policy that is asymptotically optimal for NUM is presented.  
The authors in \cite{Paschos_2019_infocom_net_util_max_exponentiated_gradient} propose an exponentiated gradient algorithm for NUM, proving its convergence rate to the optimum UA.  
In \cite{Hajek_1990_ToIT_load_balancing_adjustements} load balancing across APs is considered. Iterative and combinatorial algorithms that perform local adjustments are presented. 
In \cite{Alanyali_1995_INFOCOM_load_balancing} the dynamic load balancing is studied by capturing the system state with fluid equations, and  an asymptotically optimal simple myopic strategy is presented. 
The authors in \cite{liakopoulos_2018_infocom_robust_user_association_ultra_dense} predict future traffic based on the traffic history by using robust optimization tools and propose an iterative UA technique that minimizes costs.

UA is seen jointly with channel assignment in  \cite{koutso_2007_ToN_joint_AP_channel_assignment} for minimizing the number of channels needed to serve users. After applying an iterative load balancing algorithm, the problem  reduces to a simple channel allocation problem.
The work \cite{Papavassiliou_1998_ToVT_wireless_capacity_channel_power} additionally considers transmission power, quantifying limits of the achievable gains. 
In \cite{karaliopoulos_ICC_workshop_2020_cache_association} and \cite{Darzanos_2020_WiOpt_cache_associate_clustering} UA is seen jointly  with content caching for cache hit ratio maximization, and low-complexity practical schemes are presented. 
The fast-converging scheme of \cite{karaliopoulos_ICC_workshop_2020_cache_association} iterates between UA and content caching, while in \cite{Darzanos_2020_WiOpt_cache_associate_clustering} users are initially clustered based on their content preferences, and then clusters are assigned to APs.  
The work \cite{Chatzieleftheriou_2018_ACM_PER__JCRA} additionally considers  content recommendation. A simple three-step scheme that sequentially performs  a preference-aware UA with service guarantees, a recommendation-aware cache placement, and an adjustment of content recommendations, reveals the gains that can be achieved when UA is considered jointly with content caching and recommendations (as introduced in \cite{chatzieleftheriou_2017_INFOCOM_JCR}). 
Despite their interesting results, works \cite{Kelly_1998_JoOR_networks_fairness_stability, deVeciana_association, Paschos_2019_infocom_net_util_max_exponentiated_gradient, Hajek_1990_ToIT_load_balancing_adjustements, koutso_2007_ToN_joint_AP_channel_assignment, Papavassiliou_1998_ToVT_wireless_capacity_channel_power, karaliopoulos_ICC_workshop_2020_cache_association, Darzanos_2020_WiOpt_cache_associate_clustering, Chatzieleftheriou_2018_ACM_PER__JCRA} consider only static UA instances, work 
\cite{Alanyali_1995_INFOCOM_load_balancing} focuses on load balancing, and work 
\cite{liakopoulos_2018_infocom_robust_user_association_ultra_dense}
performs complex computations on the historical traffic.

\textbf{OCO theory.} 
The goal in OCO is the minimization of regret against a static benchmark, where regret is the worst-case deviation of the preformance of online algorithms from the optimal algorithm that knows all data in hindsight, but is restricted to a single action for the entire time horizon $T$. 
The following works consider convex and Lipschitz-continuous objective functions, adversarial constraints and decisions taken over a convex set.  
In \cite{Zinkevich_2003_ICML_online_convex_opt_gradient_ascent} a general class of Online Gradient Ascent (OGA) algorithms with $O(\sqrt{T})$ regret is introduced. 
The authors in \cite{Hazan_2012_ICML_projection_free_OL}  substitute OGA's projection with a Frank-Wolfe linear optimization step, achieving $O(\sqrt{T})$ regret for stochastic and adversarial costs. 
In \cite{Neely_2017_NIPS_OCO_Stoch_constr} time-varying stochastic constraints under 
a stochastic Slater assumption are studied, 
and a drift-plus-penalty algorithm with $O(\sqrt{T})$ expected regret is presented. 
In \cite{Liako_2019_ICML_Online_opt_regret_min_bubget_constraints} regret is systematically balanced with constraint violation. 
Combining stochastic optimization  \cite{Neely_2017_NIPS_OCO_Stoch_constr} 
and standard OCO \cite{Zinkevich_2003_ICML_online_convex_opt_gradient_ascent} methods, $O( KT/V + \sqrt{T})$ regret for $O(\sqrt{VT})$ constraint violation is achieved, where $K=T^k$, $k \in [0,1)$ and $V \in [K,T)$. 
These works do not consider the dimension of the problem in their solutions, which in our case is the size of the network, 
and either consider no constraints \cite{Zinkevich_2003_ICML_online_convex_opt_gradient_ascent}, 
or rely on  heavier assumptions on the input \cite{Neely_2017_NIPS_OCO_Stoch_constr, Liako_2019_ICML_Online_opt_regret_min_bubget_constraints}.

\textbf{OCO  in network resource allocation. } 
The authors in \cite{paschos_2020_ToN_OCO_caching} study online content caching under unknown file popularity. Their no-regret algorithm adapts caching and routing decisions to any file request pattern. 
In \cite{karagkioules_2019_OCO_streaming} an asymptotically optimal online learning algorithm for video rate adaptation in HTTP Adaptive Streaming  under no channel model assumptions is presented. 
The work \cite{Chen_2017_IEEETrans_OCO_proactive_network_resource_allocation} studies network power and bandwidth allocation under adversarial costs with bounded variations in consecutive slots. Constraints are satisfied on average, tolerating instantaneous violations. 
Under an additional Slater assumption, their algorithm achieves sublinear regret against a benchmark that takes the optimal decision in each time slot.

Our work is the first one that applies OCO to the minimum-cost UA problem. 
Our scheme achieves no-regret in UA decisions, with sublinear depencence both on the time horizon and on the network size, under no assumptions on the input.
Our work also introduces a novel periodic benchmark that generalizes state-of-the-art.

 \section{System model and problem formulation}\label{sec:model}

\begin{figure}[t!]
\centering
\includegraphics[width = 0.9\textwidth]{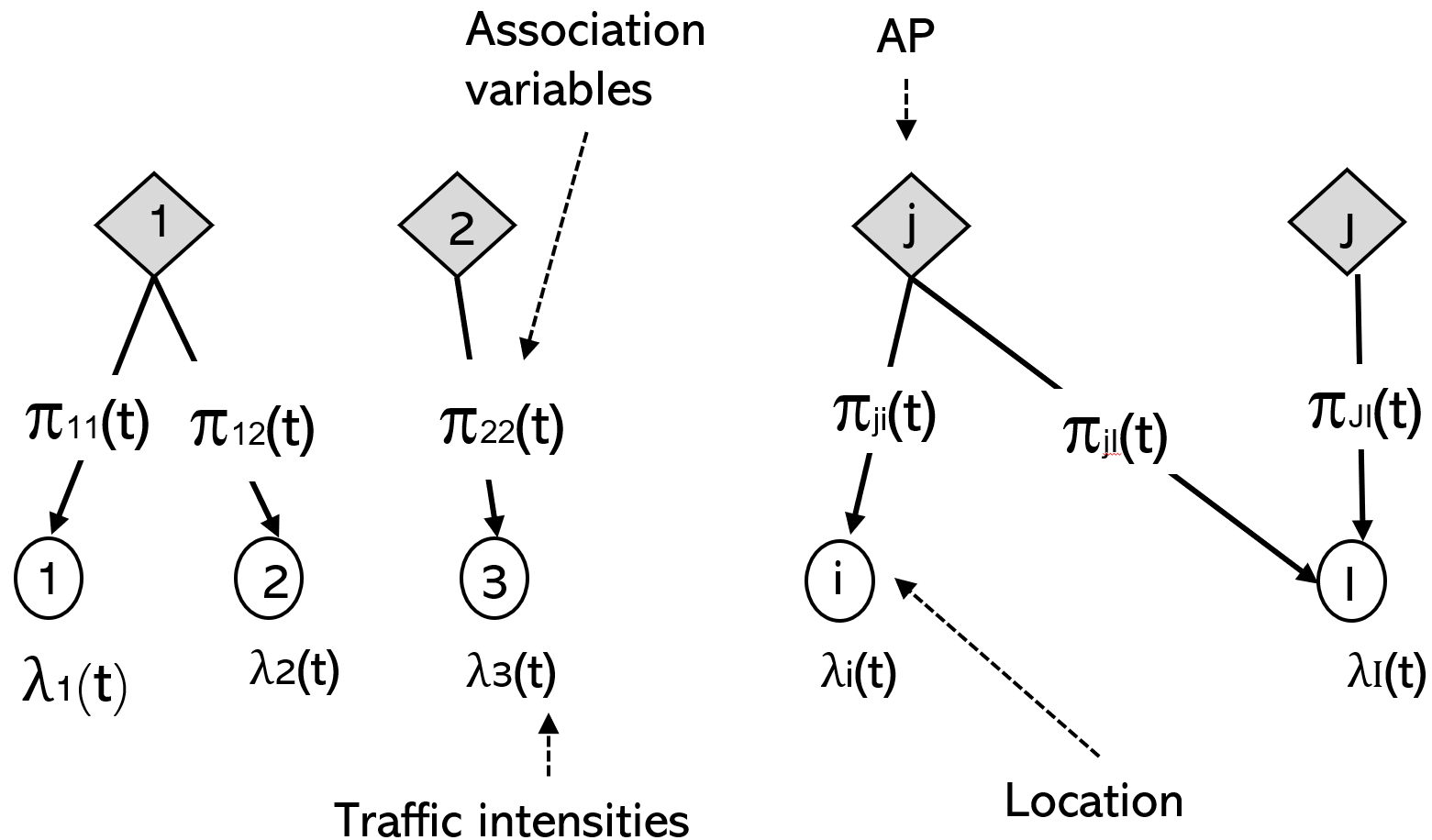}
\caption{   
At each time slot $t$ location $i\in\Ic$ requests traffic with intensity $\lambda _i(t)$, which can be split into portions $\pi_{ji}(t)\lambda_i$ and served by different APs $j\in\Nc^i$ in its neighbourhood. 
}
    \label{fig:system_model}
\end{figure}

\textbf{Basic definitions.} We start by providing some definitions and function properties that are needed throughout the paper. Although we later consider differentiable cost functions, the results of this paper are valid for any other cost function, considering $\nabla f(\xv)$ to also stand for a subgradient of $f(\cdot)$ at point $\xv$.

$\bullet$ \textit{Convexity}. A function $f(\xv):A\rightarrow B$ is convex  \textit{iff} 
$\forall\xv_1, \xv_2$ $\in$ $A,$
$$ {f(\xv_1) - f(\xv_2)
\leq 
\langle \nabla f(\xv_1),  \xv_1-\xv_2 \rangle,}$$
for $\langle \av , \bv \rangle$ the inner product of $\av$ and $\bv$.
If it exists, the Hessian matrix of a convex function is positive semi-definite, and vice versa.

$\bullet$ \textit{$p$-norm and its dual norm}. 
Let $\xv\in\R^d$. Its $p$-norm is defined as  $$\|\xv\|_p:=\left(\sum_{i=1}^{d}|x_i|^{p} \right)^{\nicefrac{1}{p}}.$$
A $q$-norm is said to be the dual of $p$-norm \textit{iff} $\frac{1}{p}+\frac{1}{q}=1.$

$\bullet$ \textit{Lipschitz-continuity}. A function 
${f(\xv):A\rightarrow B}$ is Lipschitz-continuous \textit{iff} the $p$-norm of the gradient is bounded, \ie if 
${\exists L:   \|\nabla f(\xv) \|_p  = L < + \infty. }$

$\bullet$ \textit{Strong convexity}. A function $f(\xv):A\rightarrow B$ is $\sigma$-strongly-convex  w.r.t. a $p$-norm  \textit{iff} $\forall\xv_1, \xv_2\in A$,
${f(\xv_1)
  \geq 
  f(\xv_2) + \<\nabla f(\xv_1),\xv_2-\xv_1 \> +\frac{\sigma}{2}\|\xv_2-\xv_1\|_p}
  $.

\textbf{Model components.} We consider downlink transmissions in a geographical area that is partitioned into locations $i\in\Ic$ and is covered by a set $\Jc$ of APs. 
We define the "neighbour-hood" set $\Nc_j, j\in\Jc, $ as the subset of locations that can be served by AP $j$. Similarly,  $\Nc^i, i\in\Ic, $ is the subset of APs that can serve traffic of location $i$. 
An overview of our model and  relevant 
notation are given in Fig.~\ref{fig:system_model} and Table \ref{table:notation}, respectively.  

\textbf{Location traffic. }  
We denote as $\lambdav=(\lambda_i)_{i\in\mathcal{I}}$ the traffic in-tensity vector. Each element $\lambda_i\geq 0$ is the aggregate amount (in packets/second) of requested traffic of all users in location $i$, modeled as a random variable from a general distribution.

\begin{table}[t]
\centering
\setlength\tabcolsep{1pt}
\begin{tabular}{|r|l||r|l|}
\hline
    \multicolumn{4}{c}{\textbf{Model}}\\
     \hline
    $\Ic$              & Set of locations $i$  
    &
    $\Jc$                & Set of APs $j$
    \\
    \hline 
    $\Nc^i$     &    Set of neighbour APs for location $i$ 
    &
    $\Nc_j$      &        Set of neighbour locations for AP  $j$
    \\
    \hline  
    $\lambda_i$  & Intensity of traffic requested at $i$
    &
    $\pi_{ji}$
    &      
    Fraction of $\lambda_i$ routed to $i$ by $j$
    \\
    \hline
    $\rho_j$ & Total load at AP $j$
    &
    $\rho_0$     &      Load threshold
    \\
    \hline
    $\Pi$    &          Probability simplex
    &
    $\Omega$  &      Feasibility set
    \\
    \hline
    $\phi_\alpha(\cdot)$  &    Cost  function
    &
    $T$      &      Time horizon
    \\
    \hline
    $t$      &      Time slot
    &
    $K$  &    Number of zones in each period
    \\
    \hline
    $\Wc_k$      &      Time window: Set of slots in zone $k$
    &
         &        
    \\
    \hline
    \hline
    \multicolumn{4}{c}{\textbf{Equivalent problem formulation and Association Algorithm}}
    \\
    \hline
    $V(\cdot)$  &    Penalty-featured costs
    & 
    $\Omega'$          &      Extended feasibility set
    \\
    \hline
    $L$  &    Lipschitz constant for $V$
    & 
    $h(\cdot)$          &      Regularization  function
    \\
    \hline
    $g(\cdot)$  &    Mirror function 
    & 
    $\Thetav$   &   Matrix with gradient information 
    \\
    \hline    
    $t^\tau_k$          &      $\tau$-th time slot in window $\Wc_k$
    &
     & 
    \\
    \hline    
\end{tabular}
\caption{Notation table}
\label{table:notation}
\end{table}

\textbf{Access Point (AP) load. }
The traffic requested by a location $i$ can be served by multiple APs, those in $\Nc^i.$ 
An association policy determines the association control variables $\pi_{ji} \in [0,1]$ denoting the fraction of traffic $\lambda_i$ which is routed from AP $j$ to location $i$. 
Each location's demand must be entirely served, so its association variables are constrained to lie in the probability simplex. Thus,  $\forall i\in\Ic,\, (\pi_{ji} )_{j \in \Jc} \in  \Pi,$ where: 
\begin{sal}
    \Pi = \big\{\xv \in [0,1]^{|\Jc|} :
    \sum_{j\in\Jc}x_{j} = 1\big\}.
    \label{constraint:probability_simplex}
\end{sal}
Following an association decision $\piv =(\pi_{ji} )_{j \in \Jc, i \in \Ic}$, AP $j$ transmits an aggregate demand intensity $\sum_{i\in\Nc_j} \lambda_i \pi_{ji} $. 
The packet transmission process at each AP is modeled as a queuing process. 
Prior work \cite{deVeciana_association} has shown that statistical multiplexing effects can be captured by modeling this queue  with processor sharing service. Assuming the packets have exponentially distributed sizes with mean $1/\omega$, and denoting as $C_{ji}$ the average transmission rate 
from BS $j$ to location $i$ (averaged over the channel statistics), the load $\rho_j$ of BS $j$ is
\vspace{-0.5mm}
\begin{sal}
    \rho_j(\piv , \lambdav )=\mathlarger{\sum}_{i\in\Nc_j} \frac{\lambda_i  \pi_{ji} }{\omega C_{ji}}.  
    \nonumber
\end{sal}
\vspace{-0.5mm}
Let $\rhov = (\rho_j)_{j \in \Jc}.$ 
The AP traffic load is a measure of the percentage of time the AP is busy  with packet transmission. 
When $\rho_j<1,$ AP $j$ is stable  in the sense that its packet transmission queue does not grow unbounded. 
Values close to 1 indicate large delays. If $\rho_j>1$, the AP queue is unstable and grows without limit. It results in infinite delays and bad user experience, and therefore must be avoided. 
To ensure stability and a high-quality service in terms of delay for the end-users, 
association decisions $\piv$ are constrained so that: 
\begin{sal}
    \rho_j(\piv , \lambdav )\leq\rho_0, \forall j\in\Jc,
    \label{eq:load_less_than_unity}    
\end{sal}
where $\rho_0\in(0,1)$ a load threshold. 
Combining \eqref{constraint:probability_simplex} and \eqref{eq:load_less_than_unity}, the feasible set for association variables is: 
\begin{sal}
    \Omega 
    = 
    \left\{ \piv\in \Pi^{|\Ic|}: \mathlarger{\sum}_{i\in\Nc_j} \small{\frac{\lambda_i  \pi_{ji} }{\omega C_{ji}}}\leq\rho_0, \forall j\in\Jc \right\}.
    \label{eq:feasible_region}
\end{sal}

\textbf{Cost function. }
Let $\phi(\piv ,\lambdav )$ be the system cost as a result of association policy $\piv $ under traffic  $\lambdav $.  
Our cost functions belong to the following family of convex and  Lipschitz-continuous in $\piv$ functions \cite{Kelly_1998_JoOR_networks_fairness_stability}, for $\rhov\leq\rhov_0$ and $\alpha\geq0$:
\begin{sal}
    \phi_{\alpha} (\piv ,\lambdav )  
    &= 
    \sum_{j\in\Jc}\phi_{\alpha}^j (\piv ,\lambdav ),
    \,
    \label{eqn:utility_functions}
    \\
    \text{ where}\,\,\,\,\,
    \phi_{\alpha}^j (\piv ,\lambdav )  
    &= 
    \begin{cases*}
    \frac{1}{\alpha - 1}
    (\small{1-\rho_j(\piv ,\lambdav )})^{1-\alpha}, & $\alpha \neq 1$\\   
    -
    \small{\log(1-\rho_j(\piv ,\lambdav ))}, & $\alpha =  1$  
    \end{cases*}.
    \label{eqn:utility_functions_per_AP}
\end{sal}
To confirm convexity in $\piv$ when $\rhov <\rhov_0$, observe that the cost functions are twice differentiable with positive second derivative, hence their Hessian matrix is positive semidefinite, which implies  convexity.
To confirm Lipschitz-continuity in $\piv$, observe that when $\rho_j <\rho_0$, then  $\forall j, i$,  $\nabla \phi_{\alpha} (\pi_{ji} ,\lambda_{ji} ) $ is bounded, and so  is the $p$-norm $\|\nabla \phi_{\alpha} (\piv ,\lambdav ) \|_p$. 
We will rely on both the convexity and the Lipschitz-continuity of the cost function to design our online user association algorithm and prove its performance guarantees. 

Different values of $\alpha$ lead to different cost functions. For example, for 
$\alpha = 0 ,$ \eqref{eqn:utility_functions} reduces to the total system load, 
${\small \phi_{0} (\piv ,\lambdav )  =}$
${\small \sum_{j\in\Jc} \rho_j(\piv ,\lambdav ).}$ 
For $\alpha = 2,$ it is 
${\small\phi_{2} (\piv ,\lambdav )  = }$
${\small \sum_{j\in\Jc} (1-\rho_j(\piv ,\lambdav ))^{-1}}$,  and 
\eqref{eqn:utility_functions} is equivalent to the average delay experienced by a typical demand flow in a stationary system under a temporal fair scheduler, \eg round robin 
\cite{deVeciana_association}.
 
\textbf{Optimal user association for known demand. }
If the traffic demand vector $\lambdav$ is known, 
the association policy that minimizes the system costs is found by solving problem: 
\begin{problem}[Optimal user association for known demand]\label{problem:offline}
\begin{sal}
    \min _{\piv}  \phi_\alpha(\piv, \lambdav),\nonumber \\
s.t.\quad \quad 
\sum_{j\in\Nc^i}\pi_{ji} = 1, &\quad  \forall i\in\Ic,
\nonumber\\
\rho_j(\piv, \lambdav)\leq \rho_0, &\quad \forall j\in\Jc. 
\nonumber
\end{sal}
\end{problem}
Problem \ref{problem:offline} is a convex minimization problem of a Lipschitz-continuous cost function, on the intersection of simplex and hyperplane constraints. Therefore, it can be solved through convex optimization methods \cite{Bertsekas_2015_book_Convex_opt_algo}. 
Such instances are already studied in literature. 
In this work we focus on online instances, where the traffic demand is unkown at the time of the decision.

\begin{figure}[t!]
\centering
\includegraphics[width = \textwidth]{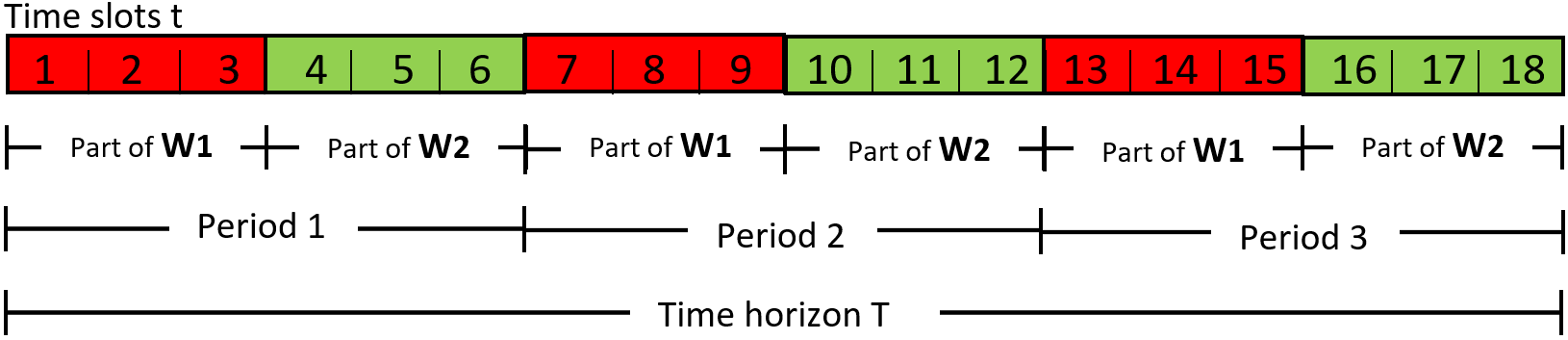}
\vspace{-2mm}
\caption{Toy example to demonstrate  periodicity, with 
$T=18$ time slots in the time horizon, 
$P=3$ periods, 
$K=2$ time zones in each period, 
$Z=3$ time slots in each time zone during each period. 
The time windows are: ${\Wc_{1}= \{1,2,3,7,8,9,13,14,15\}}$,  ${\Wc_{2}= \{4,5,6, 10,11,12, 16,17,18\}}$.
}
\label{fig:model_periodicity}
\end{figure}

\textbf{Time dynamics.} 
We capture time dynamics by denoting as $\lambdav(t),$ $\piv(t),$ and $\phi_\alpha\big(\piv(t),\lambdav(t)\big),$  the traffic vector,  association decision, and resulting system cost during time slot $t\in\{1, 2, \ldots, T\}$, where $T$ a time horizon.

\section{A novel periodic benchmark and Regret}\label{sec:regret}

\textbf{Adversarial Online Learning. }
In realistic conditions, the traffic $\lambdav(t)$ for the next time slot is unknown. Therefore, the association decisions $\piv(t)$ for slot $t$ will be computed based on knowledge of traffic demands $\lambdav(t-1).$ 
After the decision $\piv(t)$ is taken, the actual demand $\lambdav(t)$ emerges, and the actual value $\phi_\alpha\big(\piv(t),\lambdav(t)\big)$ of the cost function is revealed. 
This lack of information during the decision $\piv(t)$ at time slot $t$ may imply additional costs, or even instability  of AP packet transmission queues, due violation of the load threshold.

An appropriate setting for such online optimization problems is Online Convex Optimization (OCO) \cite{Shwartz_2012_FoundationsML_OL_OCO, mertikopoulos_2018_arxiv_tutorial_online_opt_no_regret}. 
We assume traffic demand vectors $\lambdav(t),$ and therefore cost functions  $\phi_\alpha\big(\piv(t),\lambdav(t)\big)$,  to be arbitrarily selected by an adversary who tweaks them without adhering to any probability distribution, aiming at obstructing our decisions.
While in reality the traffic vectors are not changed by such an adversary, this framework offers a convenient way to design algorithms with provable worst-case guarantees under arbitrary variations of system parameters.

\textbf{Traffic periodicity. }
The inherent nature of human activity results in traffic periodicity. 
For example, daily and weekly patterns can be observed due to people going at work or returning at home. 
Motivated by this we introduce our periodic benchmark. It generalizes state-of-the-art and characterizes the performance of online algorithms, being in between of the two extremes: the static benchmark \cite{Zinkevich_2003_ICML_online_convex_opt_gradient_ascent} and the dynamic one. 

We divide the time horizon $T$ in $P$ periods, and each period $p$ in $K$ time zones. Without loss of generality, let each time zone $k$ contain $Z$ time slots in each period. 
Then $P=\nicefrac{T}{KZ}$, and each period $p$ contains $KZ$ slots. 
This naturally defines a time window $\Wc_k$ for each time zone $k$, which includes all time slots that belong to time zone $k$, across all periods. It is: 
\begin{sal}
    \Wc_k = \Big\{ t:\,\,
    &t=KZ(p-1) + Z(k-1) + \tau ,
    \mbox{ where }
    \nonumber
    \\
    &\, \tau\in\{1, ..., L\}, 
    p\in\{1, ..., P\}
    \Big\}, \,\,k=1, ..., K.  
\nonumber
\end{sal}
The $K$ time zones define the manner in which each period is partitioned, while time windows include all time slots of a time zone across the time horizon. 
This partitioning of the time horizon captures any type of periodicity, \eg daily, weekly, or any underlying combination. 
Assuming daily periodicity in our toy example of Fig. \ref{fig:model_periodicity}, the time horizon $T$ is divided in $P=3$ days, each having $K=2$ zones. During each period, each zone contains $Z=3$ time slots, each of 4 hours duration. 

We want to stress that \textit{we do not consider periodicity on the traffic demands}: traffic vectors $\lambdav$ are considered to have \textit{arbitrary variations} during time. On the contrary, we aim to capture possible (approximate) periodic-like  “patterns” or “trends” that may exist. 
In fact, a key contribution of this work is the introduction of the following periodic benchmark.

\begin{algorithm}[t]
\caption{Optimal Periodic Static (OPS) benchmark policy}
\begin{algorithmic}[1]
\renewcommand{\algorithmicrequire}{\textbf{Input:}}
\renewcommand{\algorithmicensure}{\textbf{Output:}}
\REQUIRE Traffic vectors $\lambdav(t),\,t\in\{1,\ldots, T\},$ partition of time horizon in time windows $\Wc_k, k\in\{1,\ldots, K\}.$ 
\ENSURE  K optimal periodic static policies ${\piv^* = \big\{\piv^*[k] \big\}}_{k}$, one for each time window  $\Wc_k$.
 \FOR {$k = 1$ to $K$}
    \STATE Compute optimal static association policy in zone $k$, 
    \begin{sal}
    \pi^*[k] = \argmin_{\piv \in \Omega}\sum_{t\in \Wc_k}
    \phi_a\big(\piv,\lambdav(t)\big)
    \label{eq:static_opt_sol}
\end{sal}
\ENDFOR
\end{algorithmic}
\label{algo:static}
\end{algorithm}

\textbf{Regret against the Optimal Periodic Static (OPS) algorithm: a novel periodic benchmark.  }
Given a sequence of traffic vectors $\lambdav(1), \lambdav(2), \dots, \lambdav(T)$ over a time horizon $T$, OPS consists in finding $K$ user association policies $\piv^*[1], \piv^*[2], \dots, \piv^*[K]$, 
\textit{one for each time zone} $k$. 
Each static policy $\pi^*[k]$ is optimal regarding only traffic loads in the respective time window $\Wc_k$, and is defined as in \eqref{eq:static_opt_sol}. 
This novel periodic benchmark exploits possible approximate traffic periodicity and allows the comparison of dynamic online policies to static association rules that change according to the general traffic characteristics in each window. 
For example, it is possible to consider two different association policies, one for peak hours and one for hours with low traffic, and compare our dynamic policy against these. 
In the toy example of fig. \ref{fig:model_periodicity}, OPS would find two static association policies: one for $\Wc_1$ and one for $\Wc_2$. 
We provide its pseudocode in Algorithm  \ref{algo:static}.

A performance metric that characterizes the learning  performance of an online algorithm is \textit{regret}: 
the difference between the performance, which in our case is the experienced cost, between an online policy and a benchmark. 
Let $\piv_A(t)$ be the decision taken by an online algorithm $A$ at slot $t$. The regret ${Reg}_A(T, K)$ of $A$ with respect to OPS, for $K$ time zones in each period over a time horizon $T$, is: 
\vspace{-0.75mm}
\begin{sal}
    {Reg}_A(T, K) := 
    \sum_{t=1}^T
        \phi_a(\piv_A(t), \lambdav(t)) 
    -
    \sum_{k=1}^K      \sum_{t\in\Wc_k} 
    \phi_a(\piv^*[k], \lambdav(t)).     
    \label{eq:regret_over_periodic}
\end{sal}

\textbf{OPS \vs~existing benchmarks and regrets.}
We remind the reader of the optimal static and optimal dynamic benchmark policies as defined in \cite{Zinkevich_2003_ICML_online_convex_opt_gradient_ascent}, which we denote as  $\piv_S^*$ and $\piv_D^*(t)$, respectively. 
The optimal static benchmark knows all traffic changes in hindsight and finds \textit{one} user association policy $\piv_S^*$ that minimizes the costs over the entire time horizon, \ie 
\vspace{-0.75mm}
\begin{sal}
    \piv^*_S := \argmin_{\piv\in \Omega} \sum_{t=1}^T
    \phi_a\big(\piv,\lambdav(t)\big).
\label{eqn:opt_static}
\end{sal}
\vspace{-0.75mm}
On the contrary, the optimal dynamic benchmark knows all traffic changes but aims at minimizing the cost functions \textit{for each time slot} $t$ and finds one user association $\piv^*_{D}(t)$: 
\vspace{-0.5mm}
\begin{sal}
    \piv^*_{D}(t) := \argmin_{\piv\in \Omega} 
    \phi_a\big(\piv,\lambdav(t)\big), \, \forall t=1, \ldots T.
\label{eqn:opt_dyn}
\end{sal}
\vspace{-1.25mm}
OPS generalizes the state-of-the-art benchmarks, and the respective regrets against them.  It is easily verifiable that:  
\begin{itemize}
    \item {For $K=1,$ then $\Wc_k=\{1, \ldots, T\}$,  $\pi^*[k] \equiv \piv^*_S$, and  \eqref{eq:regret_over_periodic} reduces to the static regret.}
    \item{For $K=T,$ then $\Wc_k=\{t=k\}$, $\pi^*[k] \equiv \piv^*_{D}(t)$, and \eqref{eq:regret_over_periodic} reduces to the dynamic regret.}
\end{itemize}

\textbf{Online learning with \textquote{no regret}.} A desirable property for the regret is to scale sublinearly with the time horizon $T$, \ie  $Reg_A(T,K) =o(T)$. 
In this case, 
\begin{sal}
    \lim_{T\rightarrow +\infty}\frac{Reg_A(T,K)}{T}=0,
    \nonumber
\end{sal}
and the online algorithm $A$ is said to have "no regret", which means that it  learns to perform as well as the benchmark asymptotically as the time horizon $T \rightarrow +\infty$. 

Another desirable feature for online algorithms is to have scalable regret. This happens when regret is also sublinear to the problem dimension $d$, which in our case equals $|\Jc|\cdot|\Ic|$. It means that by increasing the size of the network by a unit, a sublinear increase in the regret is implied. At the moment, most regret results arrive at $\sqrt{dT}$. In low dimensions, this is a good result, implying the ability to learn quickly. However, as $d$ starts to grow and becomes $d \sim T$, the above expression results in a regret $O(T)$, which means that learning is not attainable in the long run. Indeed,  large systems may require a very large horizon $T$ to learn - unless we are able to decrease the dependence of regret expression to $d$.
The above is thus a property of vital importance for the envisioned future large-scaled communication networks.

\newpage
\section{Online User Association Algorithm\\ with No Regret}\label{sec:algo}

\subsection{Augmented penalty function}
\label{sec:problem}
In order to avoid overloading cells, we reformulate the user association problem with the use of a penalty function that is added to the total cost, while removing the constraints. 
The penalty is active and adds to the cost when constraints are violated, \ie when $\exists j: \rho_j>\rhov_0$.
The set of optimal solutions remains the same, because the structure of the problem and the coupling with the load-constraints now appear in the objective. 

Our penalty function $B_j(\piv (t), \lambdav(t))$ for overloading AP $j$ could be any convex and Lipschitz-continuous in $\rhov$ function, such as 
$$B_j(\piv (t), \lambdav(t)) =\psi \nabla \phi_\alpha^j(\rho_0)\cdot (\rho_j-\rho_0),$$ 
where $\psi>0$ a penalty factor for AP-overloading.
This captures the cost for each overloaded AP as the linear extension of the cost function at the overloading point $\rho_0$. 
Then, the cost function becomes: 
\begin{sal}
    V\big(\piv (t), \lambdav(t)\big) 
    & 
    = 
    \sum_{j\in\Jc} V_j\big(\piv (t), \lambdav(t)\big), \mbox{ where} 
    \label{eq:objective_OCO}
    \\
    V_j\big(\piv (t), \lambdav(t)\big) 
    & 
    =
\begin{cases}
\phi_\alpha^j(\rho_j), &\text{$\rho_j\leq\rho_0$}\\
\phi_\alpha^j(\rho_0) + 
\nabla \phi_\alpha^j(\rho_0)\cdot (\rho_j-\rho_0),\,\, &\text{$\rho_j > \rho_0$}
\end{cases}.
\nonumber
\end{sal}
The optimal user association problem reduces to: 
\begin{problem}[Online user association for unknown demand]  
\label{problem:online}
\vspace{-1mm}
\begin{sal}
    \min _{\piv(t)\in \Omega'}  
    \sum_{t=1}^T
    V\big(\piv(t), \lambdav(t)\big)
    - 
    \sum_{k=1}^K      \sum_{t\in\Wc_k} 
    V\big(\piv^*[k], \lambdav(t)\big).
    \nonumber
\end{sal}
where 
\vspace{-1.5mm}
\begin{sal}
    \Omega' &= \left\{ \piv: \piv \in \Pi^{|\Ic|}\right\}. 
    \label{set:Omega_tonos}
\end{sal}
\end{problem}
\vspace{-1mm}
This is a typical formulation for online learning problems. It aims at finding online a sequence of association policies $\piv(t),$ $t=1, ..., T$ that minimize regret, \ie the deviation of online decisions from those of an offline benchmark, which in this work is OPS. 
Compared to Problem \ref{problem:offline}, the feasibility set is expanded to a simplex for each location $i$ and it remains convex. This penalty formulation will enable us to perform a customised modification of a traditional algorithm, based on the specific characteristics of the new feasible space. 
The objective function remains convex and Lipschitz-continuous, as the sum of such functions. 
Both convexity and Lipschitz-continuity are crucial properties for proving that the online algorithm we will design has no regret against OPS.

We will analyze the regret with respect to this augmented cost. 
Since the linear part comes into play only when $\rho_j > \rho_0$ (which does not happen in the benchmark), a sublinear regret here implies sublinear regret for the $\phi_\alpha(\cdot)$ functions as well.

\begin{algorithm}[t]
\caption{Online Mirror Descent (OMD) }
\begin{algorithmic}[1]
\renewcommand{\algorithmicrequire}{\textbf{Input:}}
\renewcommand{\algorithmicensure}{\textbf{Output:}}
\REQUIRE Mirror function $g:\R^{|\Jc||\Ic|}\rightarrow \Omega'$, stepsize $\eta$, objective function $V(\cdot)$
\ENSURE User association $\piv(t), \forall t=1...T$
\STATE \textbf{Initialize:} $\Thetav(1)=\mathbf{0}$
\FOR{t=1, 2, ..., T}
\STATE decide association $\piv(t) = g\big(\Thetav(t)\big)$
\STATE update $\Thetav(t+1) = \Thetav(t) - \nabla V\big(\piv(t), \lambdav(t)\big)$
\ENDFOR
\end{algorithmic}
\label{algo:OMD}
\end{algorithm}

\subsection{PerOnE: Online user association with no regret}

\textbf{Online Mirror Descent. }
A general class of online schemes with no regret against the static benchmark is Online Mirror Descent (OMD) \cite{Shwartz_2012_FoundationsML_OL_OCO}, presented in Algorithm \ref{algo:OMD}.
It gives the opportunity to exploit the feasibility set of our problem, and leads to decision updates that lie in the feasible set without the need for expensive projections. 
OMD computes the current decision from the previous one using a simple gradient update rule. Let $\Thetav(t)$ be a matrix of dimension $|\Jc|\mbox{x}|\Ic|,$ initialized\footnote{Since the first available traffic vector is $\lambdav(1)$, the first update in \eqref{eq:theta_update} cannot be performed for $t<2. $ Thus, the initialization is performed for $t=1,$ instead of the common choice $t=0.$ } as $\Thetav(1)=\mathbf{0},$ and updated as
\begin{sal}
    \Thetav(t) = \Thetav(t-1) - \nabla V\big(\piv(t-1), \lambdav(t-1)\big), \, t>1.
    \label{eq:theta_update}
\end{sal}
During slot $t$ it is given as input to a "link" function $g(\cdot)$, that combines it with the previous decision $\piv(t-1)$ and "mirrors" it to a feasible association decision $\piv(t).$ 
More specifically, the updated user association is $\piv(t) =g(\Thetav(t)),$ where 
\begin{sal}
    g(\Thetav)
    :=
    \argmin _{\piv \in \Omega'} 
    &
    \left\{
    h(\piv)
    - 
    \<\eta\Thetav, \, 
    \piv\>
    \right\},
    \label{eq:general_mirror_function}
\end{sal}
with $\eta$ a stepsize, and $h(\cdot)$ a "regularization" function that is strongly-convex with respect to a norm over the feasible set $\Omega'$, where $\Omega'$ as in \eqref{set:Omega_tonos}.

\textbf{Regularization function.} The regularization function ensures stability of the decision and, if chosen appropriately, it leads to solutions that exploit the geometry of the problem, do not need expensive projections to the feasible space, and enjoy the no-regret property.

In our setting, we aim at finding associations that lie in the unit simplex for each location. 
Thus, each association policy is basically a set of probability distributions, one for each location. 
Since the feasibility set regarding location $i$ is the probability simplex, the most natural regularization function would be the Gibbs-Shannon entropy, 
    $$h_i(\piv) = \sum_{j\in\Jc}\pi_{ji} \log\pi_{ji},  $$
which would give the well known Exponentiated Gradient Descend (EGD). 
Here we consider the regularization function
\begin{sal}
    h(\piv) := 
    \sum_{i\in\Ic}h_i(\piv)
    =
    \sum_{i\in\Ic}\sum_{j\in\Jc}\pi_{ji} \log\pi_{ji},
    \label{eq:modified_entropic_DGF}
\end{sal}
which for a given user association policy equals the aggregate 
entropy of the associations for all locations. In Appendix \ref{appendix:lemma:modif_entropic_is_strongly_convex} we prove that:
\vspace{-3mm}
\begin{lemma}
The modified entropic regularization function in \eqref{eq:modified_entropic_DGF} is $|\Ic|$-strongly convex w.r.t the 1-norm. 
\label{lemma:modif_entropic_is_strongly_convex}
\end{lemma}

\textbf{Normalized exponentiated gradient.}
Combining \eqref{eq:modified_entropic_DGF} and \eqref{eq:general_mirror_function}, we get
\begin{sal}
    g(\Thetav) 
    =
    \argmin _{\piv \in \Omega'} 
    \left\{
    \sum_{i\in\Ic}\sum_{j\in\Jc}\pi_{ji} \log\pi_{ji},
    - 
    \<\eta\Thetav, \, \piv\>
    \right\}.
    \nonumber
\end{sal}
By differentiating with respect to $\pi_{ji}$, we get: 
\begin{sal}
    \frac
    {\partial g(\Thetav)}
    {\partial\pi_{ji}}
    & = 
    \log\left(\pi_{ji}\right) - \eta\Theta_{ji}+1,    \nonumber
\end{sal}
where $\Theta_{ji}$ is the element of matrix $\Theta$ related to AP $j$ and location $i$. 
This becomes zero at $\pi_{ji} = e^{\eta\Theta_{ji}-1}$.  
In order to ensure that the updated association variables $\pi_{ji}$ will lie in the unit simplex for each location $i$, we need to normalize the association of each location. 
Each element $\Theta_{ji}$  is thus "mirrored" through the exponentiated mirror function to: 
\begin{sal}
    g_{ji}(\Thetav)
    =
    \frac
    {e^{\eta\Theta_{ji}}}
    {\sum\limits_{j\in\Jc}e^{\eta\Theta_{ji}}}.
    \label{eq:mirror_element}
\end{sal}
Each association variable $\pi_{ji}$ is then updated through this mirroring as: 
\begin{sal}
    \pi_{ji}(t+1) 
    &= 
    g_{ji}\big(\Thetav(t+1)\big) 
    \stackrel
    {\eqref{eq:mirror_element}}
    {=}
    \frac
    {e^{\eta\Theta_{ji}(t+1)}}
    {\sum\limits_{j\in\Jc}e^{\eta\Theta_{ji}(t+1)}}
    \nonumber
    \\
    &\stackrel
    {\eqref{eq:theta_update}}
    {=}
    \frac
    {e^{\eta\Theta_{ji}(t)}e^{-\eta\nabla V\big(\pi_{ji}(t), \lambda_{i}(t)\big)}}
    {\sum\limits_{j\in\Jc}e^{\eta\Theta_{ji}(t)}e^{-\eta\nabla V\big(\pi_{ji}(t), \lambda_{i}(t)\big)}}
    \cdot
    \frac
    {\sum\limits_{j\in\Jc}e^{\eta\Theta_{ji}(t)}}
    {\sum\limits_{j\in\Jc}e^{\eta\Theta_{ji}(t)}}
    \nonumber
    \\
    &\stackrel
    {\eqref{eq:mirror_element}}
    {=}
    \frac
    {
    \pi_{ji}(t) 
    e^{-\eta \nabla V\big(\pi_{ji}(t), \lambda_{i}(t)\big)}
    }
    {
    \sum\limits_{j\in\Jc}
    \pi_{ji}(t) 
    e^{-\eta \nabla V\big(\pi_{ji}(t), \lambda_{i}(t)\big)}
    }.
    \label{eq:exp_update}
\end{sal}
This mapping is a simple normalization of the product of the previous association, multiplied with a negative exponentiation of the gradient of the objective function in the  previous step. 
The controller, thus, needs only the value of $\nabla V(\cdot)$ in order to decide the association of all locations $i$ to their neighbourhood APs $j\in\Nc^i$.
Using an adequate (for the geometry of the problem) normalization function leads us to decision updates that are always on the feasible set, avoiding expensive projections that would be otherwise necessary but prohibitive for large-scale networks.

\textbf{PerOnE: Our PERiodic, ONline, Exponentiated gradient association algorithm with "no regret".}
We design it based on the normalized exponentiated gradient-based association update \eqref{eq:exp_update}. 
We refer to it as PerOnE and we provide its pseudocode in Algorithm \ref{algo:online_EGD}. 
PerOnE exploits possible traffic periodicity and operates in each time window $\Wc_k, k=1, \ldots, K$ separately. 

Let $t_{k}^1$, $t_{k}^{\tau}$ and $t_{k}^{|\Wc_k|}$ be the first, the $\tau$-th, and last time
slot in time window $\Wc_k,$ respectively. 
For the first 
slot $t_{k}^1$ in each 
window $\Wc_k$, PerOnE does not have a previous user association $\piv$ to rely on, nor any prior information about traffic vectors $\lambdav(t)$ for $t\in\Wc_k$. Therefore, it simply splits the requested traffic evenly across neighbouring APs. 
At time $t=t_{k}^{\tau+1},$ it updates the association variables $\pi_{ji}(t_{k}^{\tau+1})$ as in \eqref{eq:exp_update}. For this update, it is based on the slot $t=t_{k}^{\tau},$ which   precedes $t_{k}^{\tau+1}$ \textit{within the same time window} $\Wc_k, \forall k$, as shown in \eqref{eqn:update_rule_EGD}.

\begin{algorithm}[t]
\caption{Periodic Online Exponentiated (PerOnE) }
\begin{algorithmic}[1]
\renewcommand{\algorithmicrequire}{\textbf{Input:}}
\renewcommand{\algorithmicensure}{\textbf{Output:}}
\REQUIRE {
Set of locations $\Ic,$ APs $\Jc$ and neighbouring APs $\Nc^i, \forall i$, 
penalty-featured cost functions $V(\cdot),$
partition of time horizon $T$ in time windows $\Wc_k,  k =1, ..., K$, 
step size $\eta.$
}
\ENSURE User association $\piv(t), \forall t=1,...,T$
\FOR{$t=1, 2, ..., T$}
\STATE {Identify time window $\Wc_k\ni t$}
\IF {$t= t_{k}^1$ for $\Wc_k$} 
    \STATE 
        Initialize association as 
        \begin{sal}
            \pi_{ji}(t_{k}^1) = 
            \begin{cases}
            \frac{1}{|\Nc^i|} & j\in\Nc^i\\
            0, & j\notin\Nc^i
            \end{cases}
        \label{eqn:EGD_initialization}
        \end{sal}
\ELSIF {$t= t_{k}^\tau$ for $\Wc_k$} 
    \STATE Update association as 
        \begin{sal}
            \pi_{ji}(t_{k}^{\tau+1}) = \frac{\pi_{ji}(t_{k}^{\tau})\cdot e^{-\eta \nabla V\big( \pi_{ji}(t_{k}^{\tau}),\lambdav_i(t_{k}^{\tau})\big)}}
            {\sum\limits_{j\in\Jc}\pi_{ji}(t_{k}^{\tau})\cdot e^{-\eta \nabla V\big(\pi_{ji}(t_{k}^{\tau}),\lambdav_i(t_{k}^{\tau})\big)}}
            \label{eqn:update_rule_EGD}
        \end{sal}
\ENDIF
\STATE Observe actual traffic $\small{\lambdav(t)}$
\STATE Compute gradient ${\small\nabla V\big( \piv(t), \lambdav(t)\big)}$
\ENDFOR
\end{algorithmic}
\label{algo:online_EGD}
\end{algorithm}

PerOnE, is a simple, projection-free and cost-efficient modi-fication of the OMD. 
It also achieves a sublinear bound on the regret against the OPS benchmark over the time-horizon $T$ and over the total number $|\Jc||\Ic|$ of decision variables.
Let 
\begin{sal}
    M_I=\max_j|\Nc_j|
    \,\, \mbox{ and } \,\, 
    M_J=\max_i|\Nc^i|
    \label{eq:max_locat_AP}
\end{sal}
be the maximum number of locations that are in range of an AP $j$ in the system, and the maximum number of APs that a location $i$ is in range of, respectively.  
Then it holds:
\begin{theorem}[PerOnE No-regret]
For a Lipschitz-continuous and convex objective function $V(\cdot),$ with $L$ Lipschitz constant, and $M_I, M_J$ as in \eqref{eq:max_locat_AP},  a stepsize $\eta$  and $T$ time horizon, 
PerOnE achieves the regret bound:
\begin{sal}
    Reg(T,K)
    \leq 
    \frac{K M_I\log(M_J)}{\eta  M_J}
    +
    \frac
    {\eta TL^2}
    {2|\Ic|}.
    \nonumber
\end{sal}
In particular, for stepsize 
$\eta = 
\sqrt{
\frac
{2K M_I|\Ic|\log(M_J)}
{TL^2M_J}
}, $
and since $M_I\leq |\Ic| $ and $\log(M_J)\leq M_J,$
we get:
\begin{sal}
    Reg(T,K)
    \leq 
    \sqrt{
    \frac
    {2K M_I T L^2\log(M_J)}
    {|\Ic| M_J }
    }
    \leq 
    L
    \sqrt{
    {2K T}
    }
    .
    \nonumber
\end{sal}
\label{theorem:periodic_regret}
\end{theorem}
Please refer to Appendix \ref{appendix:theorem:periodic_regret} for a proof. 

\textbf{Remark 1.} The EGD, obtained as the OMD with regularization function the entropic $h_i(\piv)$ for only one location, has a regret of $\sqrt{T}$ on the horizon, 
and a $\log(|\Jc||\Ic|)$ dependence on the number of association variables \cite{Shwartz_2012_FoundationsML_OL_OCO}. 

\textbf{Remark 2.} Our entropic function $h(\piv)$ that considers multiple locations, and the initialization step in \eqref{eqn:EGD_initialization}, imply a dependence of regret on topological characteristics such as the maximum number of locations $M_I$ that an AP has in its range, and the maximum number $M_J$ of APs in whose range the location belongs. Overall, its regret is sublinear on the total number of association decision variables $|\Jc||\Ic|$. 
Moreover, in realistic systems, the impact of the linear dependence in $M_I$ and that of the logarithmic dependence in $M_J$, on regret is very limited. In fact, their values can be considered constant compared to the system's dimension $|\Jc||\Ic|$, due to the progressively decreasing range of APs as technology evolves, which results in $\Nc^i$ and $\Nc_j$ being progressively smaller sets.  

{PerOnE's} regret follows the $\sqrt{T}$ dependence of EGD, and it also depends on the number $K$ of time zones. For $K=o(T),$ the regret is sublinear to the time horizon $T$, \ie 
$\lim_{T\rightarrow + \infty}  \frac{Reg(T,K)}{T} = 0$, which means that PerOnE learns association policies that are asymptotically optimal. The standard \cite{Zinkevich_2003_ICML_online_convex_opt_gradient_ascent} static regret is obtained for $K=1$, and aligns with the above.
For $K=O(T),$ the regret scales linearly with time, which is aligned with the impossibility result stated in \cite{Cover_1965_impossibility_result}: 
when the adversary can change its decision in each time slot, no-regret is not-attainable without other assumptions on the input. Indeed, PerOnE will just play the initialization for each $t$, or any other linear scaling, since it will play very few rounds for each part of the horizon. 
The state-of-the-art dynamic regret, obtained for $K=T$, aligns with the above.

\begin{table}[]
    \scriptsize
    \centering
    \setlength\tabcolsep{1pt}
        \begin{tabular}{|c||c|c|c||c|c|c|}
        \hline
            Scheme & \multicolumn{3}{c||}{$\rho_0=1$} & \multicolumn{3}{c|}{$\rho_0=0.5$}
        \\
        \hline        \hline
        &  $K=24$ & $K=12$ & $K=2$ & $K=24$ & $K=12$ & $K=2$ 
        \\
        \hline
        OPS &  234 & 936 & 2808 & 4212 & 3744 & 2808 
        \\
        \hline
        PerOnE & 127 & 57 & 9 & 145 & 72 & 11 
        \\
        \hline
        Optimum (OPS, $K=T$) & \multicolumn{3}{c||}{0} & \multicolumn{3}{c|}{0}
        \\
        \hline
        PerOnE ($K=1$) & \multicolumn{3}{c||}{1} & \multicolumn{3}{c|}{3} 
        \\
        \hline
        \end{tabular}
        \caption{Aggregate constraint violations during time horizon,\\ for different values of zones $K$ and load thresholds $\rho_0.$}
        \label{tab:violations}
    \end{table}

\begin{figure}
    \centering
    \includegraphics[clip, trim=3.5cm 9.5cm 3.5cm 9cm, width=\textwidth ]
    {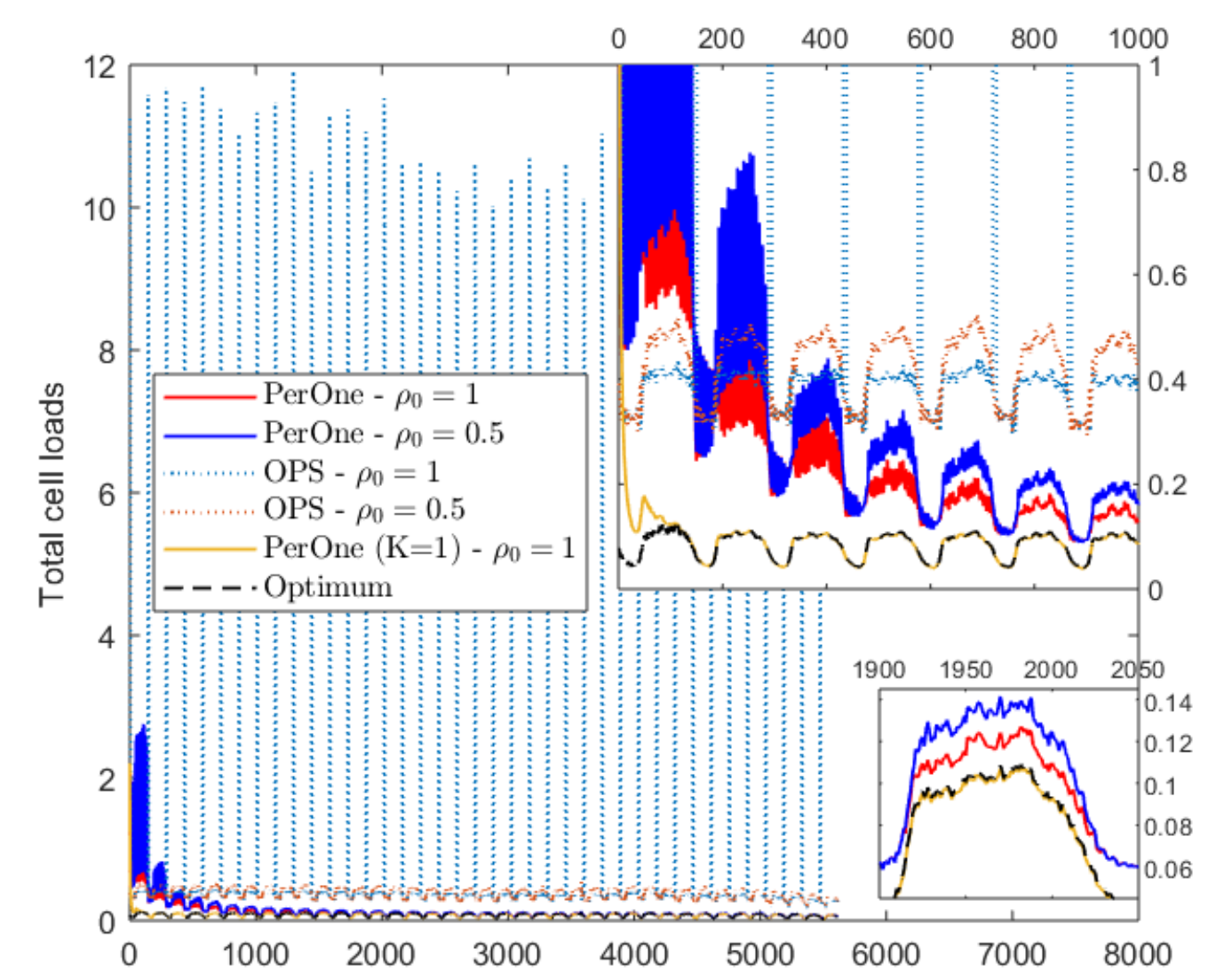}
    \caption{Total cell loads \vs~time, for $\alpha =0$ and cost function $\phi_a (\rhov)= \sum_{j\in\Jc}\rho_j$ and for load threshold $\rho_0=1$ and $\rho_0=0.5,$ and for $K=24.$}
    \label{subfig:cost_vs_r_K=24}
\end{figure}

\begin{figure}
    \centering
        \includegraphics[clip, trim=5cm 0cm 5cm 0cm, width=\textwidth]{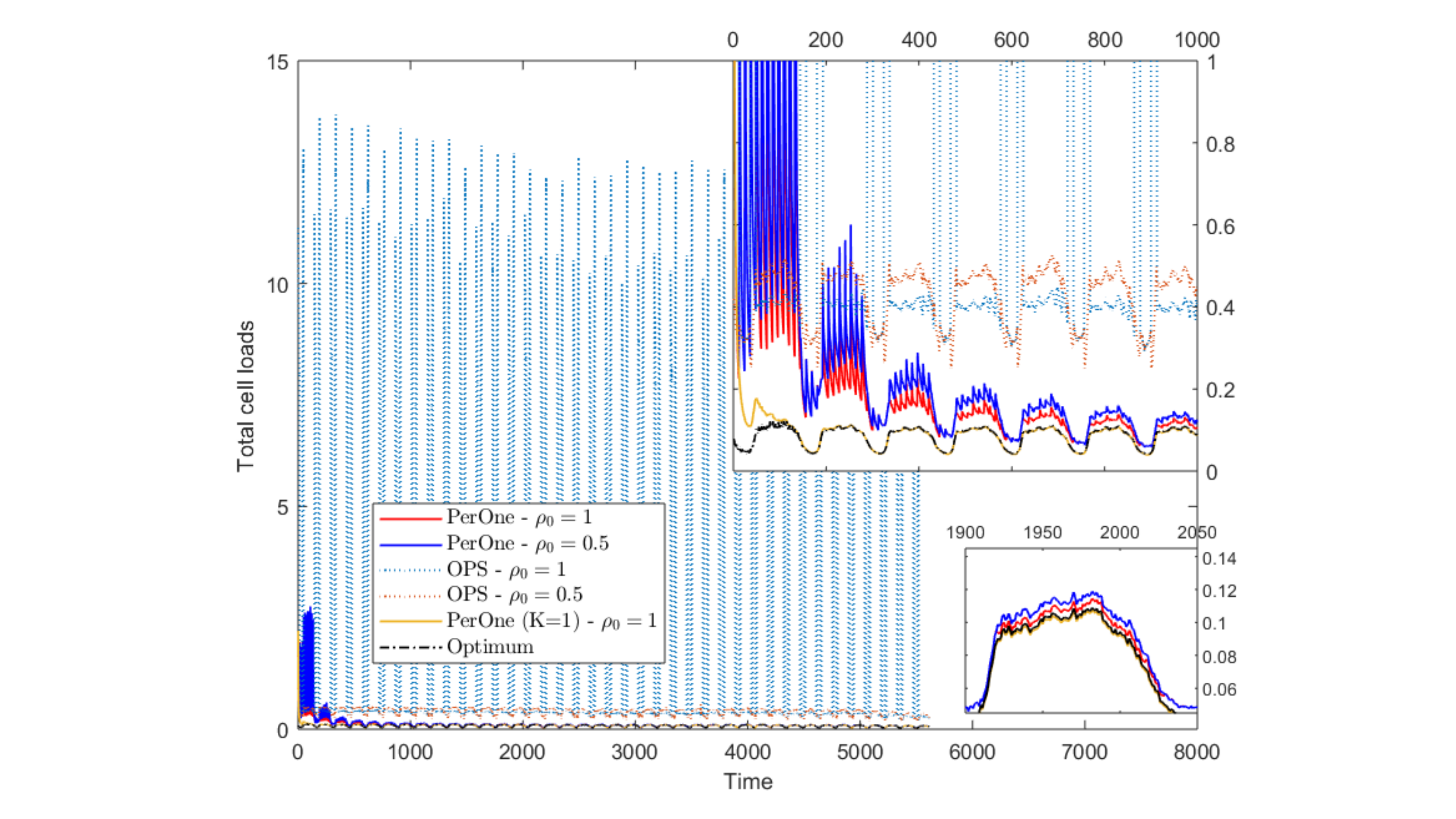}
    \caption{Total cell loads \vs~time, for $\alpha =0$ and cost function $\phi_a (\rhov)= \sum_{j\in\Jc}\rho_j$ and for load threshold $\rho_0=1$ and $\rho_0=0.5,$ and for $K=12.$}
    \label{subfig:cost_vs_r_K=12}
\end{figure}

\begin{figure}
    \centering
    \includegraphics[clip, trim=5cm 0cm 5cm 0cm, width=\textwidth]{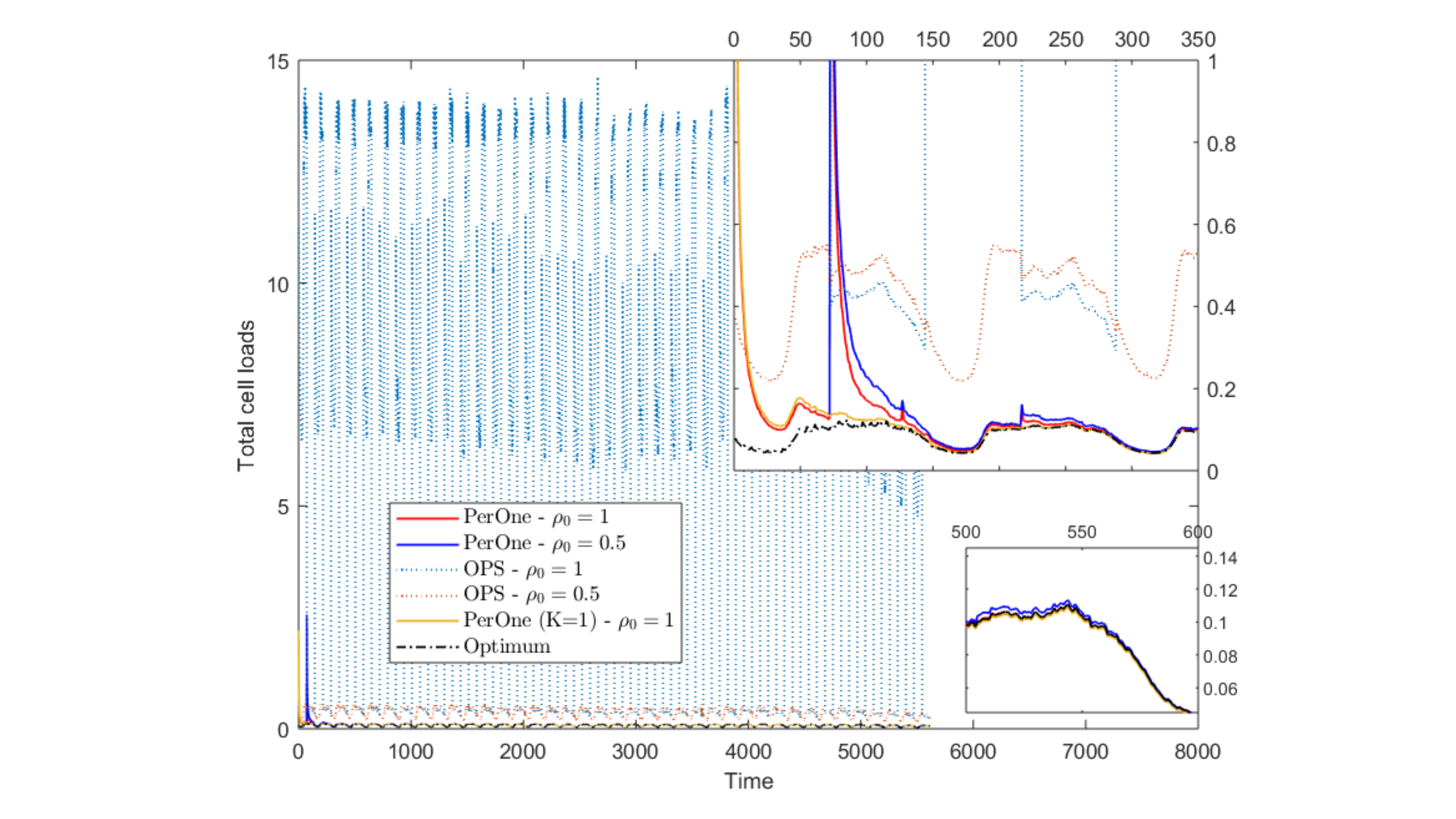}
    \caption{Total cell loads \vs~time, for $\alpha =0$ and cost function $\phi_a (\rhov)= \sum_{j\in\Jc}\rho_j$ and for load threshold $\rho_0=1$ and $\rho_0=0.5,$ and for $K=2.$}
    \label{subfig:cost_vs_r_K=2}
\end{figure}

\begin{figure}
    \centering
    \includegraphics[width=0.9\textwidth]{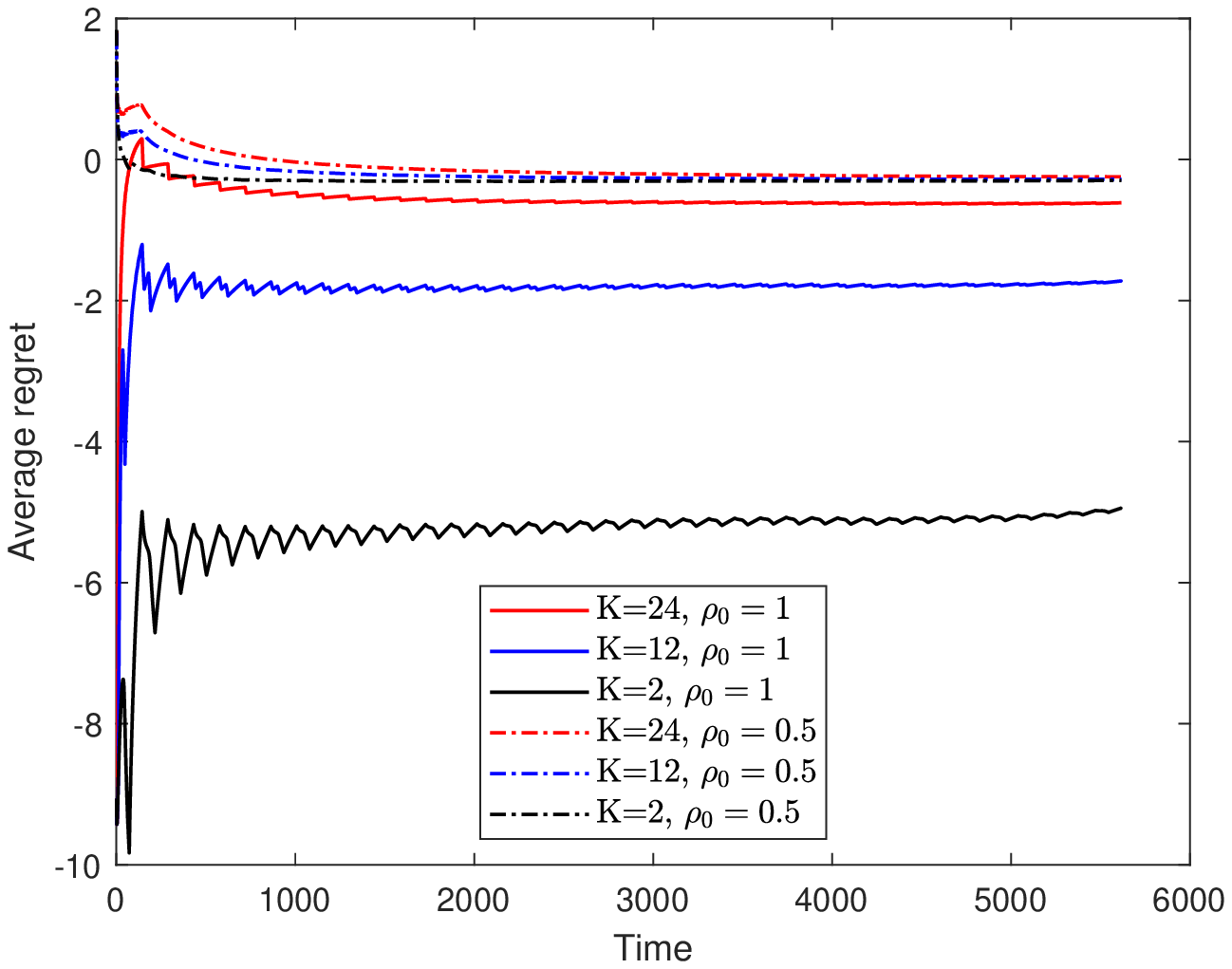}
    \caption{PerOnE's regret over OPS, for $\alpha =0$ and cost function $\phi_a (\rhov)= \sum_{j\in\Jc}\rho_j$ and for different values of $\rho_0$ and $K$.}
    \label{subfig:regret}
\end{figure}

\section{Numerical Evaluation}\label{sec:simulations}

\subsection{System architecture and traffic demand}

We perform our evaluation on the internet traffic activity of the publicly available dataset \cite{milano_dataset}. 
It provides the demand of Telecom Italia's customers in Milano, Italy, from 1/11/2013 to 1/1/2014. 
The spatial distribution $\lambda_i$ of telecommunication events is aggregated in a 100 x 100 grid of locations $i\in\Ic$. 
The temporal distribution of events is aggregated over 10-minute time intervals. 
For our analysis we consider only working days, in order to evaluate the system under high traffic and under the periodicity created by the work-cycles behaviour of people. The used dataset consists of $P=39$ days, each containing 144 time-slots, with a horizon of $T=5616$ traffic observations.

Our network architecture consists of 40 BS, most of them being close to the city center, where the load is higher. We follow the setup of \cite{liakopoulos_2018_infocom_robust_user_association_ultra_dense}, and  consider Macro- and Micro- BSs  transmitting at $P_M=43$dBm and $P_m=33$dBm, respectively. The system bandwidth is $W=10 $MHz, while the noise density is $N_0=-174$dBm/MHz. The path loss exponent is $P_{lo}=3, $ and $G_{ji}$ is the resulting coefficient for the signal degradation from AP $j$ to location $i$. 
Then, the transmission rates $C_{ji}$ between  location $i$ and AP $j$ are given by the Shannon formula: 
\begin{sal}
C_{ji} = W \log_2\left(1 + \frac{G_{ji}P_j}{W N_0 + \sum_{k\neq j}G_{ki}P_k}\right).
\nonumber
\end{sal}
We are interested in evaluating the total cell loads that arise from user association policies produced by PerOnE, and to compare them to those of policies produced by OPS. 

\subsection{Results}

We conduct a sensitivity analysis on the number $K$ of time zones and load threshold $\rho_0$ values, to capture the scenario where the maximum available resources are considered ($\rho_0=1$), and a scenario with more limited resources ($\rho_0=0.5$).  
The "Optimum" is for OPS when $K=T$ and $\rho_0=1,$ \ie it is the optimal association decision for each individual time slot $t$ under the maximum amount of resources that could be considered. The PerOnE for $K=1$ considers only one time zone, \ie runs taking as input the association policy of the previous slot, and without considering any division in the time horizon. 
For convenience, in our plots 
we provide an enlargement of the first time slots and of some slots that are indicative of how close PerOnE performs to the Optimum, at the top-right and bottom-right corner of the subfigures, respectively. 
We list some of our observations:

    \textbf{PerOnE quickly learns the optimal user association.} 
PerOnE exploits the geometry of the problem and rapidly learns the optimal user association, despite the lack of actual traffic information. 
From Figs. \ref{subfig:cost_vs_r_K=24}-\ref{subfig:cost_vs_r_K=2}, 
and Table \ref{tab:violations} we see that as the number $K$ of time zones increases, PerOnE needs more slots in order to learn not to violate constraints, to converge to optimal solutions and to produce more cost-efficient associations. 
This interesting feature allows PerOnE's solution updates to adapt to any traffic fluctuation. 
It is due to the fact that, as the number of the considered time zones decreases, the time slots that PerOnE initializes its decisions as for the expensive uniform solutions in \eqref{eqn:EGD_initialization} decreases too, similarly impacting the total cell loads.  
The contrary holds for OPS, whose static solutions benefit from a partition of the time horizon in more zones. Observe that it produces the minimum-cost static policies for a given partition of the time horizon, which does not necessarily imply that it will not have any constraint violations. In fact, as $K$ grows, OPS violates constraints during more time slots and under more limited resources. However, observe from Figs. \ref{subfig:cost_vs_r_K=24}-\ref{subfig:cost_vs_r_K=2}, that the actual cost of the produced associations grows as $K$ decreases. 

    \textbf{PerOnE effectively adapts to traffic changes. }
Despite the arbitrary and large traffic variations,  
PerOnE manages to adapt its solutions and decide cost-effective and near-optimal policies, both under high and low load threshold, as seen from Figs.  \ref{subfig:cost_vs_r_K=24}-\ref{subfig:cost_vs_r_K=2}. 
From these and table \ref{tab:violations}, it can be observed that OPS fails to adapt, thus resulting to association policies that lead to a higher system load and constraint violations.

    \textbf{PerOnE has no regret against OPS.}
Despite the large flunctuations during the duration of the day, and the lack of actual information during the decision, PerOnE manages to produce asymptotically optimal solutions, under different  partitions of each period in zones, and under different load-thresholds. As fig. \ref{subfig:regret} suggests, PerOnE's  advantage  over OPS grows as the load threshold $\rho_0$ increases and as the number $K$ of time zones decreases. Intuitively, for larger $\rho_0$ has greater "margins" to adapt to the upcoming actual traffic. Also, for smaller $K$ OPS is more restricted in its decisions, increasing PerOnE's advantage of adjusting its decisions dynamically.  

\section{Conclusions}
We assume arbitrary traffic variations over time. 
We introduce OPS, a novel periodic benchmark for online learning problems, which is significant to compare against in cases of conjectured traffic periodicity and generalizes state-of-the-art. 
We propose PerOnE, an asymptotically optimal online algorithm that produces association policies by performing a simple update.  
PerOnE learns to adapt to traffic fluctuations even under lack of actual information.
We demonstrate its no-regret property against OPS
both analytically and by performing simulations over a real-trace dataset.
Moreover, PerOnE operates under no assumptions over traffic, which renders it a great user-association option for the highly dynamic environments envisioned for the large-scaled 5G and B5G/6G networks.  
In our future work we are interested to explore algorithms that jointly learn several dynamic  parameters, for example user association and power control.

\section{Acknowledgments}
This work was supported by the CHIST-ERA LeadingEdge project, call on "Smart Distribution of Computing in Dynamic Networks" (SDCDN).

\appendices

\section{Proof of Lemma \ref{lemma:modif_entropic_is_strongly_convex}}\label{appendix:lemma:modif_entropic_is_strongly_convex}
The entropic function $h_i(\piv)=\sum_{j\in\Jc}\pi_{ji} \log\pi_{ji} $ is 1-strongly convex with respect to the 1-norm \cite{Shwartz_2012_FoundationsML_OL_OCO}, \ie 
\begin{sal}
  h_i(\piv_1)\geq h_i(\piv_2) + \<\nabla h_i(\piv_1),\piv_2-\piv_1 \> +\frac{1}{2}\|\piv_2-\piv_1\|_1, \,\forall i\in\Ic.
  \nonumber
\end{sal}
Summing over all locations $i\in\Ic$ we obtain:
\begin{sal}
  h(\piv_1)
  &\geq 
  h(\piv_2) + \sum_{i\in\Ic}\<\nabla h_i(\piv_1),\piv_2-\piv_1 \> +\frac{|\Ic|}{2}\|\piv_2-\piv_1\|_1,
  \nonumber
\end{sal}  
which due to the interchangeability of the sum and the dot-product, and due to \eqref{eq:modified_entropic_DGF}, becomes:
\begin{sal}
  h(\piv_1)
  &\geq 
  h(\piv_2) + \<\nabla h(\piv_1),\piv_2-\piv_1 \> +\frac{|\Ic|}{2}\|\piv_2-\piv_1\|_1.
  \nonumber
\end{sal}


\section{Proof of Theorem  \ref{theorem:periodic_regret}}\label{appendix:theorem:periodic_regret}

For simplicity of notation, let ${\small \zv(t):=\nabla V\big(\piv(t), \lambdav(t)\big)}$. 
Theorem 2.21 in \cite{Shwartz_2012_FoundationsML_OL_OCO} states that when the regularization function $h(\cdot)$ is $|\Ic|$-strongly-convex w.r.t. a $p$-norm and the OMD is run with mirror function as in \eqref{eq:general_mirror_function}, then: 
\begin{sal}
    \sum_{t=1}^T
    \< \zv(t),  \piv(t)- \piv^*\>
    \leq
    \frac{h(\piv^*) }{\eta}
    -
    \frac{h\big(\piv(1)\big)}{\eta}
    +
    \sum_{t=1}^T
    \frac
    {\eta \|\zv(t) \|_q^2}
    {2|\Ic|}
    ,
    \label{theorem.2.21_shwartz}
\end{sal}
where $\piv^*$ an optimal static decision over $T$ time slots as in \eqref{eqn:opt_static}, and the $q$-norm is the dual norm of the $p$-norm. 
We adopt this result and modify it to fit in the context of our periodic benchmark, using it for each time window $\Wc_k, $ separately. Thus:
\begin{sal}
\footnotesize{
    \sum_{t\in\Wc_k}
    \< \zv(t),  \piv(t)- \piv^*[k]\>
    \leq
    \frac{h(\piv^*[k])}{\eta} 
    - 
    \frac{h\big(\piv(t^{1}_k)\big)}{\eta}
    +
    \sum_{t\in\Wc_k}
    \frac
    {\eta \|\zv(t) \|_q^2}
    {2|\Ic|}
    .}\nonumber
\end{sal}
From the Lipschitz-continuity of the objective function, it exists a positive constant $L\geq \| \zv(t) \|_q, $ for all $q$-norms and time windows $\Wc_k.$ 
Thus, for $k=1, ..., K,$ it is:
\begin{sal}
    \sum_{t\in\Wc_k}
    \< \zv(t),  \piv(t)- \piv^*[k]\>
    \leq
    \frac{h(\piv^*[k])}{\eta} 
    - 
    \frac{h\big(\piv(t^{1}_k)\big)}{\eta}
    +
    \sum_{t\in\Wc_k}
    \frac
    {\eta L^2}
    {2|\Ic|}
    ,
    \label{theorem.2.21_shwartz:adapted_to_windows}
\end{sal}
Since the association variables for each location belong in $[0,1], $ for each location $i$ it holds that 
$ h_i(\piv)=\sum_{j\in\Jc}\pi_{ji} \log\pi_{ji} \leq 0$, which implies that 
\begin{sal}
    h(\piv^*[k])
    = 
    \sum_{i\in\Ic}h_i(\piv^*[k])\leq 0.  
    \label{theorem_proof:1} 
\end{sal}
Moreover:
\begin{sal}
    h\big(\piv(t^{1}_k)\big)
    &\stackrel
    {\eqref{eq:modified_entropic_DGF}, \eqref{eqn:EGD_initialization}}
    {=}
    \sum_{j\in\Jc}\sum_{i\in\Ic}
    \frac{1}{|\Nc^i|}
    \log(\frac{1}{|\Nc^i|})
    \nonumber
    \\
    &= 
    - \sum_{j\in\Jc}\sum_{i\in\Nc_j}
    \frac{\log(|\Nc^i|)}{|\Nc^i|}
    \geq
    -\frac{M_I\log(M_J)}{M_J}, 
    \, \forall k,
    \label{theorem_proof:2} 
\end{sal}
where the inequality is due to \eqref{eq:max_locat_AP}, because $M_I\leq |\Ic|$ and $M_J\leq |\Jc|.$
Then, \eqref{theorem.2.21_shwartz:adapted_to_windows} together  with \eqref{theorem_proof:1} and \eqref{theorem_proof:2} leads to:
\begin{sal}
    \sum_{t\in\Wc_k}
    \< \zv(t),  \piv(t)- \piv^*[k]\>
    \leq\,\,
    \frac{M_I\log(M_J)}{\eta M_J}
    +
    |\Wc_k|
    \frac
    {\eta L^2}
    {2 |\Ic|}, \forall k
    .
    \label{theorem_proof:3}
\end{sal}
Convexity  of the objective function implies: \begin{sal}
    V\big(\piv(t), \lambdav(t)\big) - V(\piv^*[k], \lambdav(t))
    \leq 
    \langle \zv(t),  \piv(t)-\piv^*[k] \rangle,
    \label{eq:convexity_implication_for_regret}
\end{sal}
for all $t\in\Wc_k, k=1, ..., K.$ 
Then: 
\begin{sal}
    Reg(T, K)
    &\stackrel
    {\eqref{eq:regret_over_periodic}}
    {=}
    \sum_{t=1}^T
        \phi_a(\piv(t), \lambdav(t))
    -
    \sum_{k=1}^K      \sum_{t\in\Wc_k} 
    \phi_a(\piv^*[k], \lambdav(t)) 
    \nonumber
    \\
    &\stackrel
    {\eqref{eq:objective_OCO}, \eqref{eq:static_opt_sol}}
    {\leq}
    \sum_{k=1}^K     
    \sum_{t\in\Wc_k} 
    \left(
    V\big(\piv(t), \lambdav(t)\big)
    -
    V\big(\piv^*[k], \lambdav(t)\big)
    \right)
    \nonumber
    \\
    &\stackrel
    {\eqref{eq:convexity_implication_for_regret}}
    {\leq}
    \sum_{k=1}^K     
    \sum_{t\in\Wc_k} 
    \langle \zv(t),  \piv(t)-\piv^*[k] \rangle
    \nonumber
    \\
    &\stackrel
    {\eqref{theorem_proof:3}}
    {\leq}
    \sum_{k=1}^K     
    \left(
    M_I
    \frac{\log(M_J)}{\eta M_J}
    +
    |\Wc_k|
    \frac
    {\eta L^2}
    {2|\Ic|}
    \right)
    \nonumber
    \\
    & =
    \frac{K M_I\log(M_J)}{\eta  M_J}
    +
    \frac
    {\eta TL^2}
    {2|\Ic|},
    \label{eq:regret:result}
\end{sal}
which concludes the first part of the proof.
The first equality is basically the definition of regret. 
From \eqref{eq:static_opt_sol} $\piv^*[k]$ minimizes costs for $t\in\Wc_k,$ and from \eqref{eq:objective_OCO} the objective $V(\cdot)$ is at least equal to costs. The first inequality comes from the observation that the penalty paid by the optimal benchmark $\piv^*[k]$ can't be greater than that paid by the online algorithm that takes decisions under lack of information. 
The second inequality comes as a result of the convexity of $V(\cdot),$ 
and the next from substituting the RHS of \eqref{theorem_proof:3} for each window $\Wc_k.$
The last equality holds because $\sum_{k=1}^K |\Wc_k| = T, $ since summing all time slots over all the time windows is equivalent with summing over the entire time horizon.

For the second part of the Theorem, it is easily verifiable that  \eqref{eq:regret:result} is minimized for  
$\eta = 
\sqrt{
\frac
{K M_I\log(M_J)2|\Ic|}
{TL^2M_J}
}. $
Substituting in \eqref{eq:regret:result}, and since 
$\log(M_J)<M_J$ and $M_I<|\Ic|$, we get:
\begin{sal}
    Reg(T,K)
    \leq 
    \sqrt{
    \frac
    {2K M_I\log(M_J)TL^2}
    {M_J |\Ic|}
    }
    \leq 
    L
    \sqrt{
    {2K T}
    }
    .
    \nonumber
\end{sal}

\bibliographystyle{IEEEtran}
\bibliography{bibliography.bib}

\end{document}